# Fundamental noise limitations to supercontinuum generation in microstructure fiber*

K. L. Corwin[1], N. R. Newbury[1], J. M. Dudley[2], S. Coen[3], S. A. Diddams[1], K. Weber[1], and R. S. Windeler[4]

[1]National Institute of Standards and Technology, 325 Broadway, Boulder, Colorado 80305

[2]Laboratoire d'Optique P. M. Duffieux, Université de Franche-Comté, 25030 Besançon, FRANCE

[3]Service d'Optique et Acoustique, Université Libre de Bruxelles, Av. F. D. Roosevelt 50, CP 194/5, B-1050 Brussels, BELGIUM

[4]OFS Laboratories, 700 Mountain Avenue, Murray Hill, New Jersey 07974

Broadband noise on supercontinuum spectra generated in microstructure fiber is shown to lead to amplitude fluctuations as large as 50% for certain input laser pulse parameters. We study this noise using both experimental measurements and numerical simulations with a generalized stochastic nonlinear Schrödinger equation, finding good quantitative agreement over a range of input pulse energies and chirp values. This noise is shown to arise from nonlinear amplification of two quantum noise inputs: the input pulse shot noise and the spontaneous Raman scattering down the fiber.

The generation of broadband supercontinuum spectra from the injection of femtosecond pulses into microstructure or tapered fibers has now been achieved by several groups[1, 2]. The supercontinuum is a remarkable light source, exhibiting both spatial and phase coherence, while simultaneously spanning the entire visible spectrum with brightness exceeding that of a light bulb by at least 5 orders of magnitude. These unique properties should make the supercontinuum an ideal tool for important applications including optical coherence tomography [3] and spectroscopy [4, 5]. Indeed, it has already led a revolution in frequency metrology, allowing the creation of optical atomic clocks with stability that exceeds the performance of the world's best microwave-based atomic clocks [6, 7]. However, a significant broadband amplitude noise on the supercontinuum has been observed to limit its stability, interfering with optical clocks [8, 9] and rendering the supercontinuum too noisy for many applications. This noise extends well beyond the frequency roll-off of any laser technical noise [10], and, depending on the input pulse parameters, can lead to 50 % temporal intensity fluctuations. While in some cases, empirical steps have been taken to reduce this noise, it is clear that a more complete understanding of its physical origin and scaling properties is essential if the supercontinuum is to be exploited to its full potential.

In this Letter, we show that the origin of this broadband noise is the nonlinear amplification of quantum fluctuations, both in the input laser light and in the Raman scattering process within the fiber. While this noise cannot be eliminated due to its fundamental origins, we identify methods of reducing its amplification through a judicious choice of input pulse parameters.  As well as their relevance to frequency metrology experiments, these results also represent a significant advance in the modeling of supercontinuum generation, as we present the first quantitative comparison between the measured noise on the supercontinuum and that predicted from stochastic numerical simulations.

Supercontinuum generation in bulk media, and in more conventional optical fibers, has been successfully described by the generalized nonlinear Schrödinger equation (NLSE) [11]. Recently, the generalized NLSE has proven similarly successful in describing supercontinuum generated in microstructure fiber [12-15]. However, the NLSE is insufficient to describe the broadband amplitude noise; a stochastic NLSE [16] must be employed.  Here, we have performed a quantitative study of the supercontinuum noise by using a stochastic NLSE model which rigorously includes quantum-limited shot noise on the injected input field as well as spontaneous Raman fluctuations via a stochastic Langevin source term [16].  Surprisingly, these small noise seeds are amplified into very large intensity fluctuations due to the inherent nonlinear processes involved in the supercontinuum generation.  This noise amplification is closely related to earlier work on continuum generation in more conventional optical fibers, where a similar increase in amplitude noise was observed.  This noise was attributed to modulation instability (MI) on the pulse envelope induced by amplified spontaneous emission (ASE) from the amplified input laser pulse [17, 18].

The overall noise spectrum of the supercontinuum from the microstructure fiber will be comprised of both broadband amplitude noise and a low-frequency component resulting from the laser technical noise. Experimental and numerical studies show the supercontinuum spectral and coherence properties to depend sensitively on initial input conditions [15, 19, 20], indicating that any technical noise on the laser will also result in large amplitude fluctuations across the supercontinuum. This low-frequency technical noise, which arises from laser power fluctuations or beam pointing stability, can be mitigated by the proper choice of experimental parameters, for example a quieter pump laser. In contrast, the broadband noise resulting from the input shot noise is fundamental to the supercontinuum generation process since the input shot noise and the spontaneous Raman scattering are quantum noise sources.

Figure 1 shows the experimental setup. An $Ar^+$ laser-pumped femtosecond titanium:sapphire laser provides pulses with a typical bandwidth of ~45 nm full width at half maximum (FWHM) centered at 810 nm at a 100 MHz repetition rate. A double-passed fused-silica prism pair introduces a linear chirp on the laser pulses, and interferometric autocorrelation measurements are used to infer both the input-pulse duration and chirp magnitude (in $fs^2$), assuming a $sech^2$ pulse intensity envelope. The chirped pulses with typical energies of 0.9 nJ are injected into a 15 cm long microstructure fiber with zero group-velocity-dispersion at 770 nm [1], and the output is characterized using an optical spectrum analyzer and an apparatus dedicated to measuring the relative intensity noise (RIN). Here, the supercontinuum is attenuated to prevent detector saturation, spectrally filtered by a monochromator with 8 nm bandwidth, and directed to either an infrared or visible detector. The resulting electrical signal is fed into an electrical spectrum analyzer, where the RF noise power above the detector noise floor is measured. The RIN in dBc/Hz is obtained from this noise power, divided by the RF electrical bandwidth and the total detected power, and is measured at 10 nm increments across the supercontinuum. To avoid contributions from laser technical noise, the noise power is measured at Fourier frequencies of 3 MHz and above, where the titanium:sapphire laser is approximately shot-noise limited [10].

The experimentally measured spectral width and the noise are shown in Fig. 2 as the solid lines. As noted by other groups, we see complicated spectral structure on the supercontinuum [12-15, 19]. The RIN measurements also reveal for the first time the dramatic and complicated wavelength-dependent structure of the supercontinuum noise, where fluctuations as high as 20 dB are common. Under a wide variety of input pulse conditions, the resulting supercontinua exhibit a consistent dip in the RIN at the input laser wavelengths ($\lambda_L \sim 810$ nm) and also across the Raman soliton on the infrared side of the spectrum ($\lambda_R \sim 1300$ nm). Otherwise, there is no universal correlation between the RIN and the optical spectrum. Figure 3 shows the experimental dependence of RIN on Fourier frequency for representative wavelengths of a typical data set. While the RIN varies greatly with wavelength, it is clearly white noise, *i.e.* independent of Fourier frequency [21].

The numerical simulations which model these measurements are based on the generalized NLSE [16, 20]:

$$\frac{\partial E(z,t)}{\partial z} = i\sum_{k\geq 2}\frac{i^k \beta_k}{k!}\frac{\partial^k E}{\partial t^k} + i\gamma\left(1+\frac{i}{\omega_0}\frac{\partial}{\partial t}\right)\left[E(z,t)\left(\int_{-\infty}^{t} R(t')|E(z,t-t')|^2 dt' + i\Gamma_R(z,t)\right)\right]$$

Here $E(z,t)$ is the complex pulse envelope in a co-moving frame, the $\beta_k$'s describe the fiber dispersion [1] and the nonlinear coefficient $\gamma = 100$ W$^{-1}$ km$^{-1}$ at 810 nm. The response function $R(t) = (1-f_R)\delta(t) + f_R h_R(t)$ includes both instantaneous and delayed Raman contributions with the fractional Raman contribution $f_R = 0.18$. For $h_R$, we used the measured Raman response of silica. Spontaneous Raman noise appears as the multiplicative stochastic variable $\Gamma_R$, which has frequency domain correlations $\langle \Gamma_R(\Omega,z)\Gamma_R^*(\Omega',z')\rangle = (2 f_R \hbar \omega_0/\gamma) |\text{Im } h_R(\Omega)| [n_{th}(|\Omega|)+U(-\Omega)] \delta(z-z') \delta(\Omega-\Omega')$ where the thermal Bose distribution $n_{th}(\Omega) = [\exp(\hbar\Omega/k_B T)-1]^{-1}$ and U is the Heaviside step function. The input pulse initial conditions are those of the experimentally measured pulse duration and chirp, with the addition of quantum-limited shot noise. We stress that the magnitude of the quantum noise terms on the input pulse and due to spontaneous Raman scattering have no adjustable parameters. The RIN is obtained from the simulations by Fourier analysis of numerically generated supercontinuum obtained over an ensemble of typically 128 simulations with different initial random noise, assuming a 100 MHz repetition rate as in the experiment.

The dashed lines of Fig. 2 show numerical results for one set of simulations. Good qualitative agreement is observed between the experimental and simulated spectra and RIN. The size of the fluctuations in RIN with wavelength, the average level of the RIN, and most features are reproduced in the simulations. In particular, the decrease in RIN around the input laser and Raman soliton wavelengths ($\lambda_L$ and $\lambda_S$, respectively) appears both in the experiment and theory.

The supercontinuum spectrum and noise level depend strongly on the input parameters of the laser pulse and many sets of data similar to those presented in Fig. 2 were taken under a variety of input conditions. The RIN always exhibits the complicated wavelength dependence shown in Fig. 2, but it is inconvenient to compare many such curves directly. Therefore, in subsequent figures only the median RIN value is given, calculated across all wavelengths for which there is sufficient optical power. The resulting median RIN indicates the overall RIN of the data set, but it should be remembered that the deviations from this median value with wavelength are substantial. In fact, the statistics of these deviations are roughly consistent with Gaussian optical intensity fluctuations with unit fractional standard deviation. Because of averaging effects, both the experimentally measured and simulated RIN depend on the spectral bandwidth of the monochromator. For both, in these results the spectral bandwidth was 8 nm. Reducing the spectral bandwidth to 1 nm increases the median RIN by ~ 3dB.

Both the spectral width and the RIN increase with input pulse energy, as shown in Fig. 4 for a moderately chirped input pulse. The values measured experimentally (triangles, solid line) are reproduced well in the simulation results (circles, dashed line). Significantly, the spectral width increases with injected pulse energy at the expense of a corresponding increase in the noise. In fact, the linear increase in the RIN (in dBc/Hz) translates to an exponential increase in the associated fractional intensity fluctuations with injected energy. Indeed, at the largest spectral width of ~600 nm, the relatively small RIN of -100 dBc/Hz corresponds to pulse-to-pulse fluctuations of ~7 %.

Although the results above appear to suggest that the broadest supercontinuum spectral widths are necessarily associated with the largest RIN, additional experiments show that precise control of the input pulse chirp permits the generation of octave-spanning supercontinua with near detection shot-noise limited RIN. Figure 5 shows experimental measurements (triangles) and numerical simulations (circles) of the supercontinuum spectral width and the median RIN as a function of input chirp over the range $-500 fs^2$ to $+600 fs^2$[22]. As expected, large supercontinuum spectral widths are observed with the shortest (near transform-limited) input pulses, because shorter pulse duration at constant pulse energy implies higher peak power and thus enhanced nonlinear spectral broadening. In contrast, the median RIN, which depends strongly and asymmetrically on the pulse chirp, is smallest at the shortest pulse durations. At large negative chirps of $\sim-400\ fs^2$, corresponding to pulse widths of ~60 fs, the RIN values can reach $-83$ dBc/Hz, corresponding to 50 % fluctuations in the pulse-to-pulse amplitude. However, for pulses that are near transform-limited or with a small positive chirp ($<+200\ fs^2$), RIN values are only -130 dBc/Hz, which is just above the detection shot-noise limit for our apparatus. Again, the measured dependence of RIN on chirp agrees well with the results of the simulation. Some scatter is observed in the data, especially for large positive and negative values of chirp. This is attributed in part to uncertainty in the pulse chirp and energy. These data are taken with a laser spectral width of 45 nm; data taken at input spectral widths of 27 nm and 55 nm show similar dependencies.

While the noise above results from both input noise seeds, simulations show that the input shot noise is the dominant noise seed, and that Raman scattering plays only a relatively minor role. When only Raman scattering noise is included numerically, the RIN is reduced by 20 dB; when only shot noise is included, the RIN is reduced by only 3 dB. It is surprising that for an input shot noise of $-172$ dBc/Hz, the noise at the output can be as large as $-80$ dBc/Hz, corresponding to a nonlinear amplification of ~90 dB. In fact, there is a strong link between this nonlinear amplification and the initial spectral broadening, as shown in Fig. 6, where the simulated evolution of the spectral width and RIN is plotted as a function of propagation distance. The majority of the spectral broadening occurs in the first ~1 cm of propagation, which is also the distance scale on which the input shot noise is most strongly amplified. The strong dependence of the final RIN on the pulse chirp observed in Fig. 5 can also be understood in the context of Fig. 6. For short transform-limited pulses or pulses with a small positive chirp, which undergo initial compression, the spectral broadening is more rapid than for pulses with a large negative chirp. The more rapidly the spectrum broadens, the more rapidly the pulse will spread temporally through fiber dispersion, leading to a reduced overall noise amplification (akin to modulation instability gain), and therefore a reduced overall RIN.

In conclusion, we have experimentally characterized the broadband noise on supercontinua in microstructure fiber. Numerical simulations using the stochastic generalized NLSE show that this broadband noise results from the very basic noise processes of amplified input shot noise and Raman scattering. Thus, the supercontinuum output can exhibit excess noise approaching 50 % amplitude fluctuations that arise directly from the shot noise on the input laser pulse. While the noise grows exponentially with input power, it is at a minimum for the shortest input pulse duration, which is the same condition that yields the widest spectrum. These conditions bring the supercontinuum closer to the ideal realization of a broadband, phase-coherent source.

We thank Brian Washburn, Sarah Gilbert, and Leo Hollberg for valuable discussions.

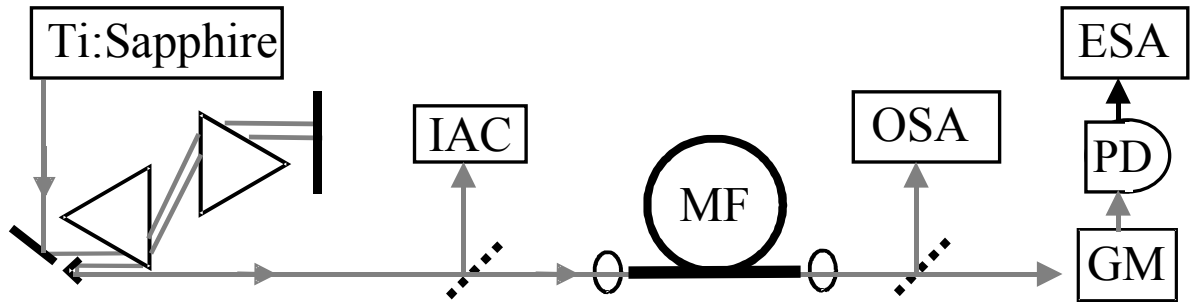

Figure 1: Simplified schematic of the experimental setup. IAC, interferometric autocorrelator; MF, microstructure fiber; OSA, optical spectrum analyzer; GM, grating-based monochromator; PD, photodiode; ESA, electrical spectrum analyzer.

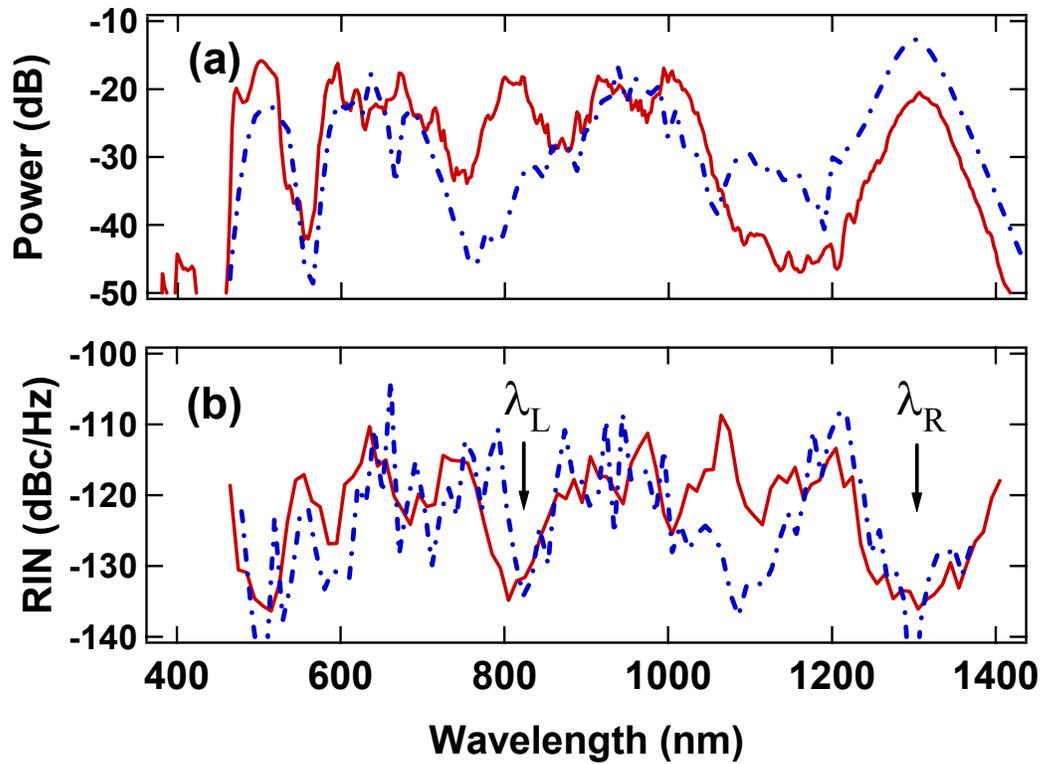

Figure 2: (a) Spectrum and (b) total RIN as a function of wavelength across the supercontinuum for experiment (solid lines) and theory (dashed line) for an input pulse duration of 22 fs FWHM and a spectra bandwidth of 45 nm FWHM (*i.e.,* with minimal chirp [22] ) at an RF Fourier frequency of 3 MHz.

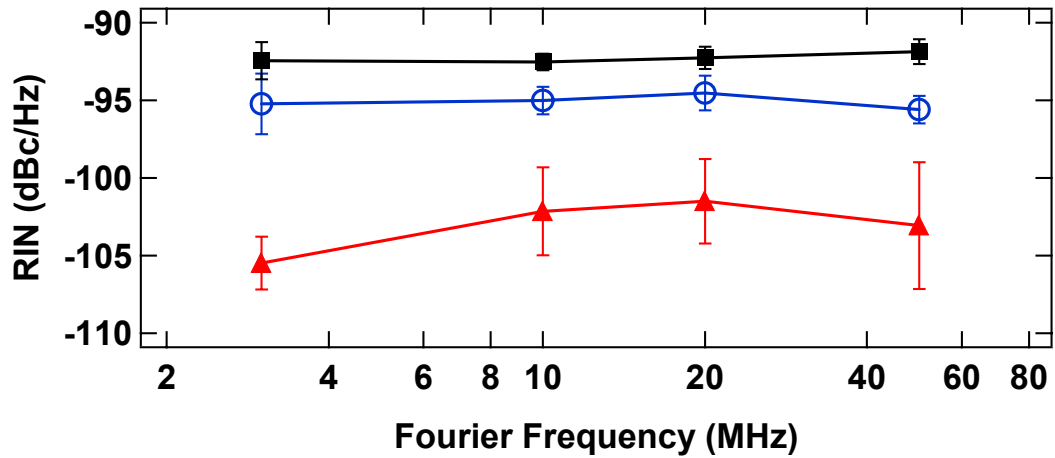

Figure 3: RIN as a function of Fourier frequency for three representative wavelengths: 620 nm (circles), 820 nm (triangles), and 920 nm (squares). The input pulse duration of 50 fs and pulse bandwidth of 45 nm corresponds to -290 $fs^2$ chirp.

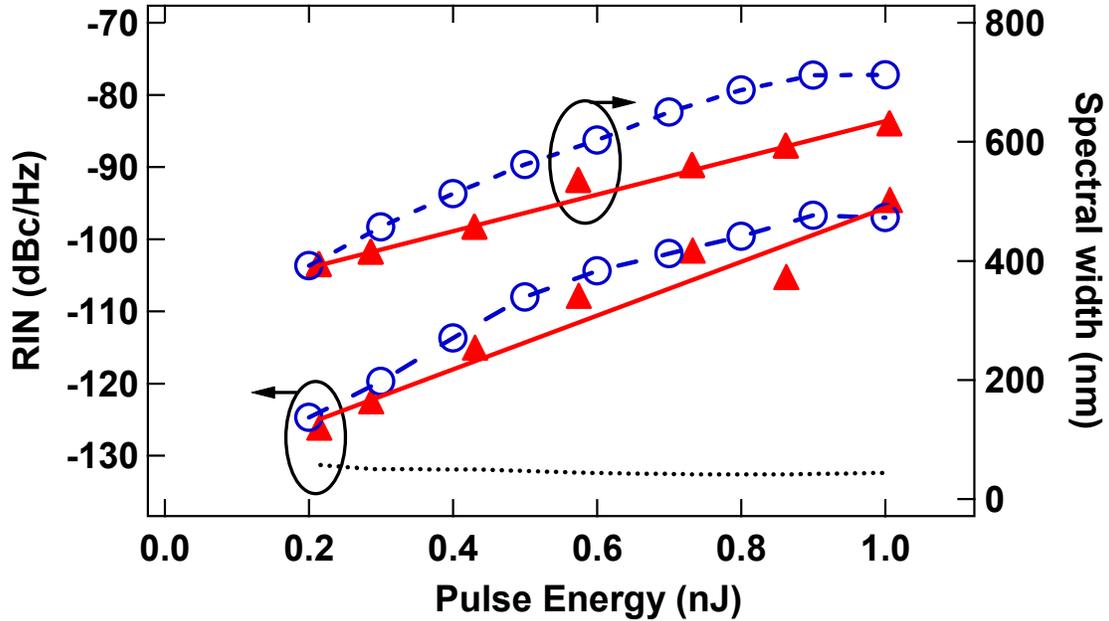

Figure 4: The RIN and corresponding –20 dB width of the supercontinuum as a function of the average power exiting the fiber for experiment (triangles) and theory (circles). Solid lines represent a linear fit to the experimental data. The dotted line is the contribution to the noise from the shot noise on the detected light. The input pulse duration of 47 fs and bandwidth of 42 nm corresponds to a chirp of -280 fs$^2$.

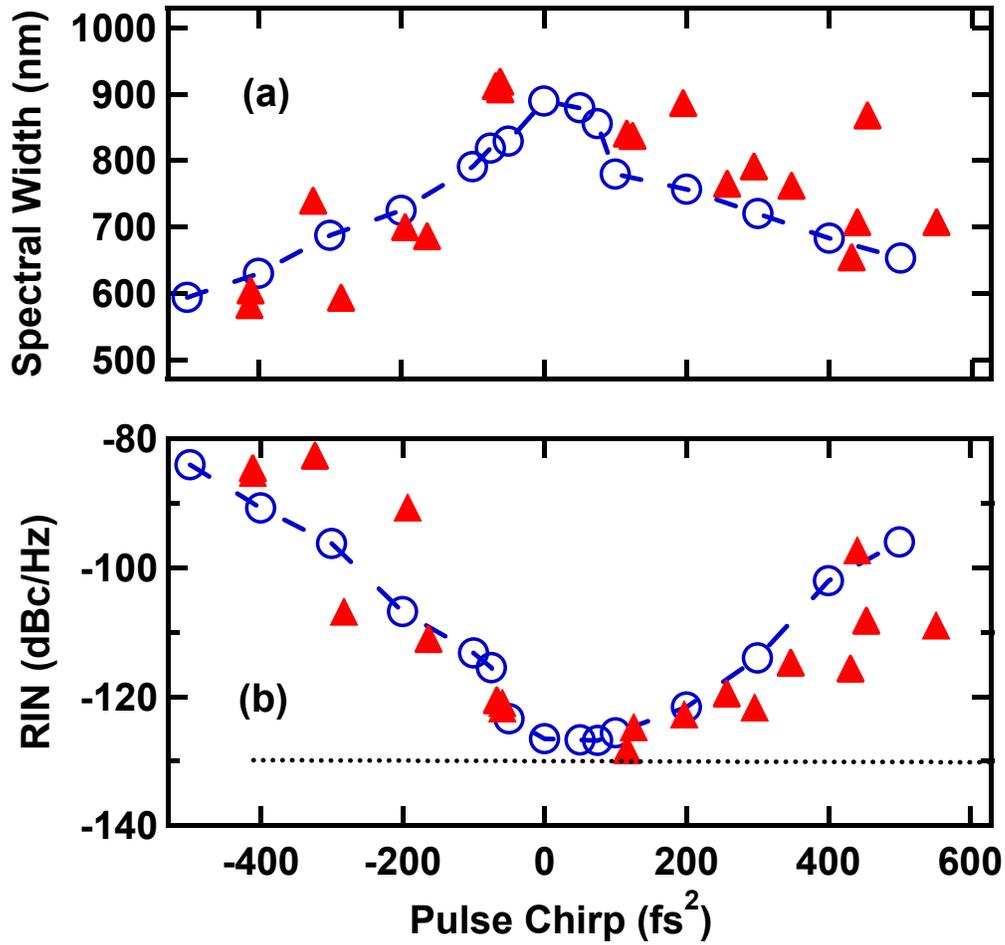

Figure 5: The median total RIN and supercontinuum spectral width as a function of pulse chirp for experiment (triangles) and theory (circles). The dotted line is the detection shot noise contribution to the total RIN. A chirp variation of 0 – 650 $fs^2$ corresponds to a range of pulse widths from ~20 to 90 fs for a pulse bandwidth of 45 nm. The uncertainty in the experimental pulse chirp is about ± 30 $fs^2$.

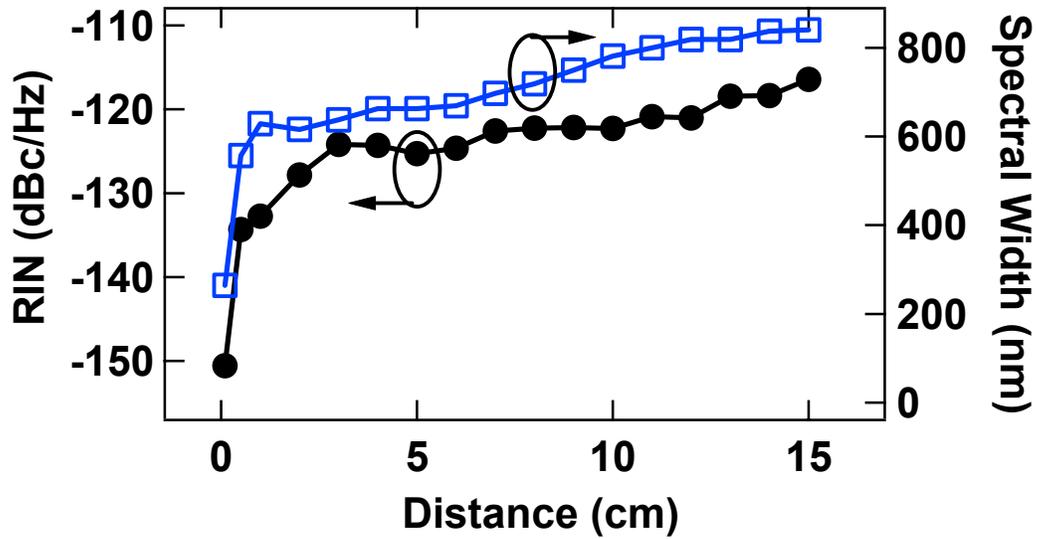

Figure 6:  Simulation results showing median RIN and –20dB spectral width as a function of propagation distance in the fiber.  The input pulse duration of 22 fs corresponds to second order dispersion of -65 $fs^2$ for a pulse bandwidth of 45 nm.  The input energy is 0.85 nJ.